**Do dogs live in joint families? Understanding allo-parental care in free-ranging dogs**

**Manabi Paul[1] and Anindita Bhadra[1,*]**

[1]Behaviour and Ecology Lab, Department of Biological Sciences, Indian Institute of Science Education and Research Kolkata, India

[*]Behaviour and Ecology Lab, Department of Biological Sciences,

Indian Institute of Science Education and Research Kolkata

Mohanpur Campus, Mohanpur,

PIN 741246, West Bengal, INDIA

*tel.* 91-33-66340000-1223

*fax* +91-33-25873020

*e-mail:* abhadra@iiserkol.ac.in

**Abstract**

Cooperative breeding is an excellent example of altruistic cooperation in social groups. Domestic dogs have evolved from cooperatively hunting and breeding ancestors, but have adapted to a facultatively social scavenging lifestyle on streets, and solitary living in human homes. Pets typically breed and reproduce under human supervision, but free-ranging dogs can provide insights into the natural breeding biology of dogs. We conducted a five year-long study on parental care of free-ranging dogs in India. We observed widespread allo-parenting by both adult males and females. Allomothers provided significantly less care that the mothers, but the putative fathers showed comparable levels of care with the mothers. However, the nature of care varied; mothers invested more effort in feeding and allogrooming, while the putative fathers played and protected more. We were unsure of the relatedness of the pups with the putative fathers, but all the allomothers were maternal relatives of the pups, which provides support for both the "benefit-of-philopatry" and "assured fitness returns" hypotheses. Free-ranging dogs are not cooperative breeders like wolves, but are more similar to communal breeders. Their breeding biology bears interesting similarities with the human joint family system.

Keywords: free-ranging dogs; maternal care; allomother; putative father; kin selection; philopatry

**Introduction**

Cooperation and conflict drive the dynamics of social groups in species as diverse as insects to humans. While selfishness is easily explained by the theory of natural selection (1,2), cooperation and altruism are more difficult to understand as behavioural traits (3). The most extreme form of cooperation in animal societies is manifested by cooperative breeding, where only a few individuals in a social group reproduce and the others help to rear their offspring while forfeiting reproduction themselves (4,5). Cooperative breeding is observed in social insects like ants, bees, wasps and termites, but is relatively less common in mammals. Among mammals, naked mole rats, lemurs and some species of canids like wolves and coyotes are well known examples of cooperative breeders (5–9).Cooperative breeding is sometimes confused with communal breeding, though these are quite distinct breeding systems. In communally breeding species, multiple females share dens or birthing sites and help each other to rear offspring (6,8). Thus the reproductive hierarchy evident in cooperative breeders is absent in communal breeders.

For the last few decades, cooperative breeding has been mostly explained by kin selection theory (10,11). It has been argued that the benefits gained by the alloparent through inclusive fitness are not enough to compensate for their own reproduction (12). Therefore kin selection is not a sufficient condition for the evolution of cooperative breeding. However, kin selection, in conjugation with philopatry, i.e., the tendency of an individual to stay in or return to a particular area, leading to an increased probability of kin living in close proximity, could explain the evolution of altruism in animal groups (13). Females of most group-living mammals are reported to be philopatric (14) and the average kinship between females is highest for smaller groups (15,16). Hence cooperative breeding could be a consequence of kin selection and philopatry in such groups; for example, subordinates of a wolf pack are usually philopatric offspring of the dominant breeding pair (17,18) that share a common territory and exhibit cooperative breeding (5).

Domestic dogs (*Canis lupus familiaris*) share a common ancestry with modern day gray wolves (*Canis lupus lupus*) (19), but show much variation in their social organization. They are capable of living solitarily as pets, in artificially formed groups as pack dogs, and as natural social groups in free-ranging populations (20–22). They have a promiscuous mating system that lacks reproductive hierarchy (23,24). Their group dynamics greatly depend on their mating and denning seasons (22), and often multiple females of a group give birth in neighbouring dens. Maternal care is the predominant form of care received by the pups (25) where mothers adjust their care with pup age and litter sizes (Paul et al. submitted). They are predominantly scavengers, surviving on human-generated wastes, but are capable of forming large packs to hunt down animals like goats and deer (26–28). Though free-ranging dogs do not show reproductive hierarchies like other cooperatively breeding canids, they do show allocare by males and related females to some extent (25,29). Allonursing is not voluntarily offered to non-filial pups, but is a manifestation of milk theft by pups from any available lactating female (Paul and Bhadra, submitted). Thus the free-ranging dogs present an interesting situation which seems to be somewhat intermediate between the strict social hierarchies observed in wolves, coyotes, wild dogs, etc. and the solitary lifestyle of canids like foxes and jackals. Alloparenting in dogs has been reported by us in a case study (29), but has never been studied in detail, and could provide interesting insights into the social system of dogs. We carried out a long term behavioural study in India to understand the patterns and implications of alloparenting behaviour in groups of free-ranging dogs.

**Methods**

In a field based study we collected behavioural data from 15 dog groups having a set of 23 mother-litter units over a period of 11 weeks, from the 3$^{rd}$ to 17$^{th}$ weeks of pup age. Each group was observed for two morning (0900-1200h) and two evening (1400-1700h) sessions spread over two week blocks. Each three hour observation session consisted of 18 scans and 18 all occurrences sessions (AOS), amounting to a total of 9108 scans of one minute each and 9108 AOS of five minutes each for all the 23 groups. We thus had a total of 910 hours of data on 50 adults and 84 pups for our analysis. The study was conducted in various areas of West Bengal, India, selected on the basis of availability of dog groups and convenience of long term observations.

Any behaviour that was shown by an adult towards a pup, which enhanced the pup's chances of survival, was designated as care (30). Any adult male that provided care in any form to the pups was designated as a "putative father" (PF), and any adult female other than the mother who provided care to the pups was designated as an allomother (AM). One of the litters (group PF$_2$) lost their mother in a car accident, just one day after their birth and their grandmother (maternal) "adopted" them and took care of them till her death in the 10$^{th}$ week of their age. Here we have considered this group as a special case and compared them with both the maternal and allomaternal care sets. Please see the Table 1 for the group details.

Care shown by the mother, putative father and allomother, were labeled as maternal care, male-allocare and female-allocare respectively. We sorted the total care shown by the adults towards the focal pups into two categories, active care and passive care. Any behaviour that involved direct interaction between the care giver and the pups, and is energy intensive was considered as active care. Behaviours that require no direct interaction with the focal pups, but can provide care in terms of protection and social bonding were considered as passive care (Paul and Bhadra, submitted).

Statistics:

We used linear mixed-effects models (LMM) in R (using "nlme" package) (31) to check the effects of pup age, and the litter size for the respective pup age on the care given by the putative father and the allomothers separately. In the models pup age and litter size have been incorporated as the fixed effects, while the male and female-allocare have been considered as the response variables. Group identity and year of observations have been included in the models as the random effects (for more details please see the ESM 1- ESM 4).

For statistical analysis, we used StatistiXL 1.10, Statistica version 12 and R statistics.

**Results**

Allocare was observed in 19 mother-litter units from 15 dog groups, of which 10 litters received all the three types of care i.e. maternal care, male-allocare and female-allocare. Four litters received male-allocare and maternal care while four others received female-allocare and maternal care. Pups of $PF_2$ group received only allocare from their grandmother and putative father (Table 1).

(i) Total care:

Pup age significantly affected the level of allocare shown by the putative father and allomother (LMM: male-allocare: t = 4.79, p <0.0001; female-allocare: t = 4.94, p <0.0001), but both forms of allocare were independent of litter size (Table 2, ESM1). Male-allocare, female-allocare and maternal care were compared considering age as the common predictor variable and no difference was evident between the maternal care and male-allocare (ANOVA: F = 11.97, p <0.0001; Post-hoc tukey test: maternal care vs male-allocare: p = 0.49), though female-allocare differed from both (Post-hoc tukey test: maternal care vs female-allocare: p < 0.0001; male-allocare vs female-allocare: p = 0.003).

(ii) Active care:

Unlike the active maternal care (Paul et al. submitted), active male and female-allocare only depended on pup age (LMM: male-allocare: t = 2.21, p = 0.03; female-allocare: t = 2.07, p = 0.04) but not on their litter size (Table 3, ESM2). Active male-allocare was comparable with both maternal care (ANOVA: F = 6.89, p <0.003; Post-hoc tukey test: active maternal care vs active male-allocare: p = 0.07) and female-allocare (Post-hoc tukey test: active female-allocare vs active male-allocare: p = 0.338) (Fig. 1a).

(a) Maternal care vs male-allocare: The mother and the putative father seemed to provide comparable levels of active care towards the focal pups. However, a closer investigation of the pre-weaning phase of pup development (3$^{rd}$ to 8$^{th}$ week) revealed interesting patterns. The mothers showed higher levels of active care than the putative fathers (T test: T = 3.82, p = 0.01) (Fig. 1a), and they budgeted their time in the various care-giving behaviours differently. For first three weeks of observations (3$^{rd}$ to 5$^{th}$ week of pup age), 76% to 86% of the active maternal care comprised of suckling and pile sleep. These behaviours are energy intensive and require constant physical contact with the pups. Play and protection replaced these behaviours as the pups grew older (Fig. 1b). In case of male-allocare, play and protection consistently contributed to (69 ± 12) % of active male-allocare, throughout the entire period of observations (fig. 1c).

(b) Maternal care vs female-allocare: In some cases, pups received female-allocare as early as their 3$^{rd}$ week of age, but the level of female-allocare was significantly lower than maternal care (2 tailed t test: t = 9.35, p < 0.0001) (Fig. 2). The level of female-allocare increased from the 3$^{rd}$ to the 9$^{th}$ week of pup age (Linear regression: $R^2$= 0.751, std. β= 0.867, p= 0.02) that again decreased as the pups grew older (Linear regression: $R^2$= 0.62, std. β= -0.787, p= 0.01) (Fig. 2). All the allomothers were related to the pups to which they provided care. Here we present the relatedness estimates considering random mating with unrelated males (Table 1).

(iii) Passive care:

Passive care remained comparable over the three types of care throughout the duration of observations (ANOVA: F = 1.94, p = 1.6) (Fig. 3), and like maternal care, both female and male-allocare only depended on pup age (LMM: male-allocare: t = 4.45, p < 0.0001; female-allocare: t = 4.74, p < 0.0001) and not on the litter size (Table 4, ESM3).

(iv) Grandmother's care:

The PF$_2$ grandmother provided significant levels of active care including allonursing to her grandpups. The care provided by her was comparable with active maternal care but not with active female-allocare (Post-hoc tukey test: active care: maternal care vs grandmother's care: p = 0.993, female-allocare vs grandmother's care: p = 0.007; suckling: maternal care vs grandmother's care: p = 0.150, female-allocare vs grandmother's care: p <0.0001) (ESM4; Fig. 4).

**Discussion**

We observed allocare by both males and females in social groups of free-ranging dogs, but such care, though quite common, was not ubiquitous. While maternal care was the predominant form of care received by pups (25, Paul et al. submitted), allocare by both males and females was equally common (25,29). Interestingly, the overall levels of care provided by the putative fathers was comparable to that of the mothers, and much higher than that provided by the allomothers. However, mothers and putative fathers differed in the nature of care that they provided to the pups. Mothers invested a substantial portion of their time activity budgets in energy demanding behaviours like suckling, piling up with pups, allogrooming, etc. that are crucial for the pups' early stage development. The putative fathers invested most of their effort in play and protection – thus mothers feed and fathers play, thereby showing division of responsibilities in pup rearing. Social play has a vital role in the juveniles' behavioural development (32) and putative fathers seems to play a significant role in this process.

Female-allocare, although not comparable to the levels of maternal care or male-allocare, allows the pups to gain some extra benefits through milk-theft (Paul and Bhadra, submitted). In spite of being unwilling to nurse the pups (Paul and Bhadra, submitted), the allomothers voluntarily performed behaviours like allogrooming, play, pile sleep and protection towards the non-filial pups, independent of their litter size but not of their age. Pups received allocare as they started to come out of their den, facilitating the increased contact with allomothers and putative fathers. This also provided pups ample opportunities for milk-theft from the allomothers, thereby creating a situation of potential conflict for the pups and their allomothers. Interestingly, all the cases of female-allocare observed were between related individuals; aunts, older sisters and grandmothers showed allocare to pups. Hence, while female-allocare could impose a cost on the allomothers, philopatry could reduce the cost by providing inclusive fitness benefits to females that provided care to related pups. In a populations that faces high early life mortality (33), any additional care received by pups could help to increase their survival probability, thus making allocaring a stable strategy in such a species. The case study of the grandmother that "adopted" her orphaned grandpups provides support to this idea.

Allocare by adults occurs in cooperatively breeding species, where adults provide care to non-filial offspring even at an expense to their own reproduction (5,9,34). In communally breeding species, allocare can provide mutual benefits to females that give birth in each other's vicinity, and such mutualistic relationships have been speculated to play a role in the evolution of cohesive social systems (35). Philopatry and kin selection can jointly provide a premise for high assured fitness returns in both cooperatively and communally breeding systems (36). Free-ranging dogs live in small social groups and have a promiscuous mating system (22,24), which allows for mingling between groups during the mating season. No reproductive hierarchy is evident within the groups (23,24), and individuals often disperse out of their natal groups (37), leading to group fission. Unlike cooperatively breeding canids like wolves and coyotes, the dogs don't usually hunt, and they tend to forage alone most often (22). Hence social cohesiveness in the dogs is not driven by the inherent need for efficient hunting, as in other social canids (38). On the contrary, scavenging as a foraging strategy is most efficient for solitary foragers (26,39). Nevertheless, we observed allocare in 82% of the total observed groups. While we could not be completely sure of the relatedness of the allocaring males with the pups, all the allocaring females were related to the pups receiving care. We have provided the relatedness estimates between the allomothers and pups considering random mating with unrelated males, but the actual relatedness values are likely to be much higher, considering the promiscuous, multiple mating system of dogs which allows mating between close kin. Thus related females denning in close proximity provided allomaternal care to pups, and this lends support to the benefits-of-philopatry hypothesis (40) and to the theory of assured fitness returns, originally proposed to explain the evolution of social behaviour in insects (36). Further genetic studies, especially to estimate relatedness values with allocaring males could provide further insights into the evolutionary dynamics of alloparental care in the dogs

Our study suggests a highly flexible nature of the breeding system in free-ranging dogs, where maternal care alone can be sufficient for the survival and development of pups, but allocare provides additional benefits to the pups. Male-allocare is relatively easily explained if the care giving males are fathers of the pups. In a promiscuously breeding species that is also philopatric, the inclusive fitness benefits can intensify for males if they breed with related females. In a competitive environment with irregular resources (26) and high early life mortality (33), providing care to non-filial pups is costly, but could provide inclusive fitness benefits to the allocaring female. Especially in the face of high mortality, such care can assure fitness benefits (36) for females in a given breeding season, thereby compensating for the cost of providing care, even when care is snatched in the form of milk-theft (Paul and Bhadra, under review). Allocare can also help to build social cohesiveness and increase cooperation within the group, and might even provide future reproductive benefits to the care-giver (3). The free-ranging dogs therefore are more similar to communal breeders like lemurs (7) than their cooperatively breeding relatives, the wolves (5). Interestingly, with a wide spectrum of possibilities including single mothers, caring female relatives, non-caring male and female group members, caring males and even adoptions, the free-ranging dog breeding biology bears great resemblance to the joint family system of humans.

**Ethics:** No dogs were harmed during this work. All work reported here was purely observation based, and did not involve handling of dogs in any manner. The methods reported in this paper were approved by the animal ethics committee of IISER Kolkata (approval number: 1385/ac/10/CPCSEA), and was in accordance with approved guidelines of animal rights regulations of the Government of India.

**Competing financial interests:** We report no competing financial interests.

**Author contributions:** MP carried out the field work, participated in data entry and carried out all the statistical analyses. MP and AB participated in the design of the study and drafted the manuscript. Both the authors gave final approval for publication.

**Acknowledgements**

We thank Ms. Sreejani Sen Majumder for her help in some of the fieldwork and Mr. Shubhra Sau for his help in part of the data entering process. We would like to thank Dr. Anjan K. Nandi for his inputs on statistical analyses.

**Funding**

This work was funded by projects from Council for Scientific and Industrial Research, India; SERB, Department of Science and Technology, India and Indian National Science Academy, India; and supported by IISER Kolkata, India.

**Tables**

| Serial no. | Year | Group name | Mother-litter group id | Litter size | Group size | Group composition | No. of AMs | Relationship between pups and AM |
|---|---|---|---|---|---|---|---|---|
| 1 | 2010-11 | CAN1 | CAN1 | 5 | 6 | MO + Pups + AM (0) + PF (0) + OJ (0) | - | - |
| 2 | 2010-11 | BUD | BUD1 | 4 | 5 | MO + Pups + AM (0) + PF (0) + OJ (0) | - | - |
| 3 | 2010-11 | LEL1 | LEL1 | 2 | 3 | MO + Pups + AM (0) + PF (0) + OJ (0) | - | - |
| 4 | 2010-11 | S1 | S1 | 2 | 3 | MO + Pups + AM (0) + PF (0) + OJ (0) | - | - |
| 5 | 2011-12 | BSF1 | RS4 | 5 | 9 | MO + Pups + AM (0) + PF (1) + OJ (2) | - | - |
| 6 | 2011-12 | PLT1 | JCB | 2 | 9 | MO + Pups + AM (1) + PF (0) + OJ (5) | 1 | Elder sister ($r = 0.25$) |
| 7 | 2011-12 | PLT1 | MDB1 | 5 | 9 | MO + Pups + AM (1) + PF (0) + OJ (2) | 1 | Grandmother ($r = 0.25$) |
| 8 | 2011-12 | CAN2 | CAN2 | 5 | 7 | MO + Pups + AM (1) + PF (0) + OJ (0) | 1 | Grandmother ($r = 0.25$) |
| 9 | 2011-12 | GH | GH2 | 6 | 8 | MO + Pups + AM (1) + PF (0) + OJ (0) | 1 | Aunt ($r = 0.125$) |
| 10 | 2011-12 | LEL2 | LEL2 | 2 | 4 | MO + Pups + AM (1) + PF (0) + OJ (0) | 1 | Grandmother ($r = 0.25$) |
| 11 | 2011-12 | S2 | S2 | 3 | 5 | MO + Pups + AM (0) + PF (1) + OJ (0) | - | - |
| 12 | 2013-14 | BSF2 | RS1 | 2 | 11 | MO + Pups + AM (1) + PF (1) + OJ (6) | 1 | Elder sister ($r = 0.25$) |
| 13 | 2013-14 | BSF2 | RS2 | 4 | 11 | MO + Pups + AM (1) + PF (1) + OJ (4) | 1 | Aunt ($r = 0.125$) |
| 14 | 2013-14 | BSF2 | RS3 | 2 | 11 | MO + Pups + AM (1) + PF (1) + OJ (6) | 1 | Aunt ($r = 0.125$) |
| 15 | 2013-14 | PF | PF1 | 5 | 8 | MO + Pups + AM (1) + PF (1) + OJ (0) | 1 | Unknown |
| 16 | 2013-14 | PF | **PF$_2$** | 6 | 9 | **Pups + AM (2) + PF (1) + OJ (0)** | 2 | Grandmother ($r = 0.25$); aunt ($r = 0.125$). |
| 17 | 2013-14 | CAN3 | CAN3 | 6 | 8 | MO + Pups + AM (0) + PF (1) + OJ (0) | - | - |
| 18 | 2013-14 | PLT2 | MDB2 | 5 | 8 | MO + Pups + AM (0) + PF (1) + OJ (1) | - | - |
| 19 | 2014-15 | BSF3 | BBR | 4 | 19 | MO + Pups + AM (4) + PF (1) + OJ (9) | 4 | Elder sister ($r = 0.25$); aunt ($r = 0.125$); grandmother ($r = 0.25$); cousin ($r = 0.0625$). |
| 20 | 2014-15 | BSF3 | KTI | 2 | 19 | MO + Pups + AM (4) + PF (1) + OJ (11) | 4 | Aunt ($r = 0.125$); grandmother ($r = 0.25$); Mother's aunt and Mother's grandmother ($r = 0.125$). |
| 21 | 2014-15 | BSF3 | WHI | 2 | 19 | MO + Pups + AM (4) + PF (1) + OJ (11) | 4 | Mother's niece and aunt ($r = 0.125$); grandmother ($r = 0.25$); cousin ($r = 0.625$). |
| 22 | 2014-15 | BSF3 | BRN | 2 | 19 | MO + Pups + AM (4) + PF (1) + OJ (11) | 4 | Aunt ($r = 0.125$); grandmother ($r = 0.25$); Mother's aunt and Mother's grandmother ($r = 0.125$). |
| 23 | 2014-15 | BSF3 | RS5 | 3 | 19 | MO + Pups + AM (4) + PF (1) + OJ (10) | 4 | 2 elder sisters ($r = 0.25$); 2 nieces ($r = 0.125$). |

**Table 1:** Table shows the group details including the identities of observed mother-litter units, litter size at birth, composition of observed groups, number of allomothers present in each group and their minimum relatedness (r) (considering random mating with unrelated males) with the focal pups. MO, OA, PF and AM have been used as the codes for the mother, other adults (excluding the MO, PF and AM), putative fathers, allomothers respectively. OJ represents the pups or juveniles other than the focal pups of the mother-litter units. $PF_2$ group (bold highlighted) lost their mother.

|  | Value | Std. Error | DF | t-value | p-value |
|---|---|---|---|---|---|
| **Putative father (PF)** |  |  |  |  |  |
| (Intercept) | 0.3345553 | 0.09900394 | 126 | 3.379212 | 0.0010 |
| **age** | **0.0270875** | **0.00565348** | **126** | **4.791293** | **0.0000** |
| LS | -0.0062323 | 0.01902409 | 126 | -0.327598 | 0.7438 |
| **Allomother (AM)** |  |  |  |  |  |
| (Intercept) | 0.21143078 | 0.11482146 | 121 | 1.841387 | 0.0680 |
| **age** | **0.03026155** | **0.00612711** | **121** | **4.938958** | **0.0000** |
| LS | -0.03117045 | 0.02175889 | 121 | -1.432539 | 0.1546 |

**Table 2:** Results of the linear mixed effects models for the effect of pup age (age) and the current litter size (LS) on the proportion of time spent by the putative fathers (PFs) and allomothers (AMs) in total allocare, shown toward the focal pups.

|  | Value | Std. Error | DF | t-value | p-value |
|---|---|---|---|---|---|
| **Putative father (PF)** | | | | | |
| (Intercept) | 0.03572216 | 0.05493874 | 126 | 0.6502181 | 0.5167 |
| age | **0.00612957** | **0.00277573** | **126** | **2.2082712** | **0.0290** |
| LS | 0.00520547 | 0.01034108 | 126 | 0.5033780 | 0.6156 |
| **Allomother (AM)** | | | | | |
| (Intercept) | 0.04242796 | 0.03867422 | 121 | 1.0970607 | 0.0680 |
| age | **0.00388757** | **0.00187833** | **121** | **2.0696973** | **0.0406** |
| LS | 0.00007415 | 0.00525762 | 121 | 0.0141041 | 0.9888 |

**Table 3:** Results of the linear mixed effects models for the effect of pup age (age) and the current litter size (LS) on the proportion of time spent by the putative fathers (PFs) and allomothers (AMs) in active allocare, shown toward the focal pups.

|  | Value | Std. Error | DF | t-value | p-value |
|---|---|---|---|---|---|
| **Putative father (PF)** | | | | | |
| (Intercept) | 0.21470119 | 0.11486552 | 126 | 1.869153 | 0.0639 |
| age | **0.02450343** | **0.00550545** | **126** | **4.450761** | **0.0000** |
| LS | 0.00442425 | 0.02004886 | 126 | 0.220673 | 0.8257 |
| **Allomother (AM)** | | | | | |
| (Intercept) | 0.16712635 | 0.10099926 | 121 | 1.654728 | 0.1006 |
| age | **0.02631783** | **0.00554810** | **121** | **4.743573** | **0.0000** |
| LS | -0.03068309 | 0.01961587 | 121 | -1.564197 | 0.1204 |

**Table 4:** Results of the linear mixed effects models for the effect of pup age (age) and the current litter size (LS) on the proportion of time spent by the putative fathers (PFs) and allomothers (AMs) in passive allocare, shown toward the focal pups.

**Figures**

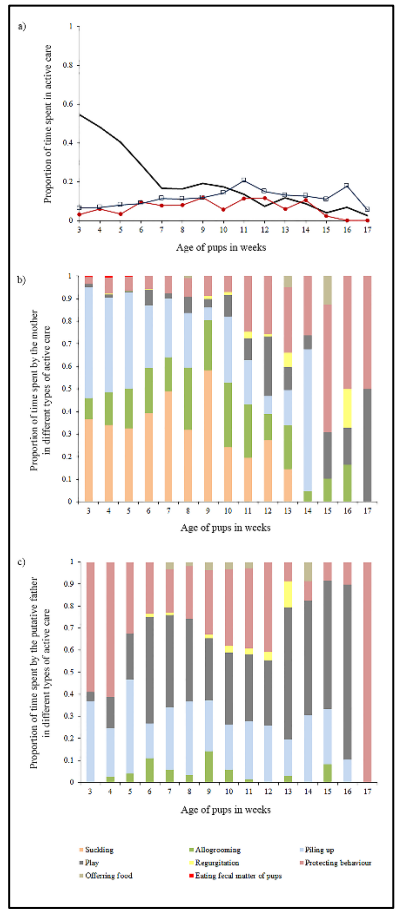

**Figure 1:** a) Line graph showing the proportion of time spent in active care by the mothers, putative fathers and allomothers. The thick black line designates the mothers, thin black line with open squares designates putative fathers and the red line with solid circles is for allomothers. b) Stacked bar diagram showing how the mothers budgeted their time in the various care-giving behaviours over pup age. c) Stacked bar diagram showing the proportion of time spent by the putative fathers in various active care over pup age.

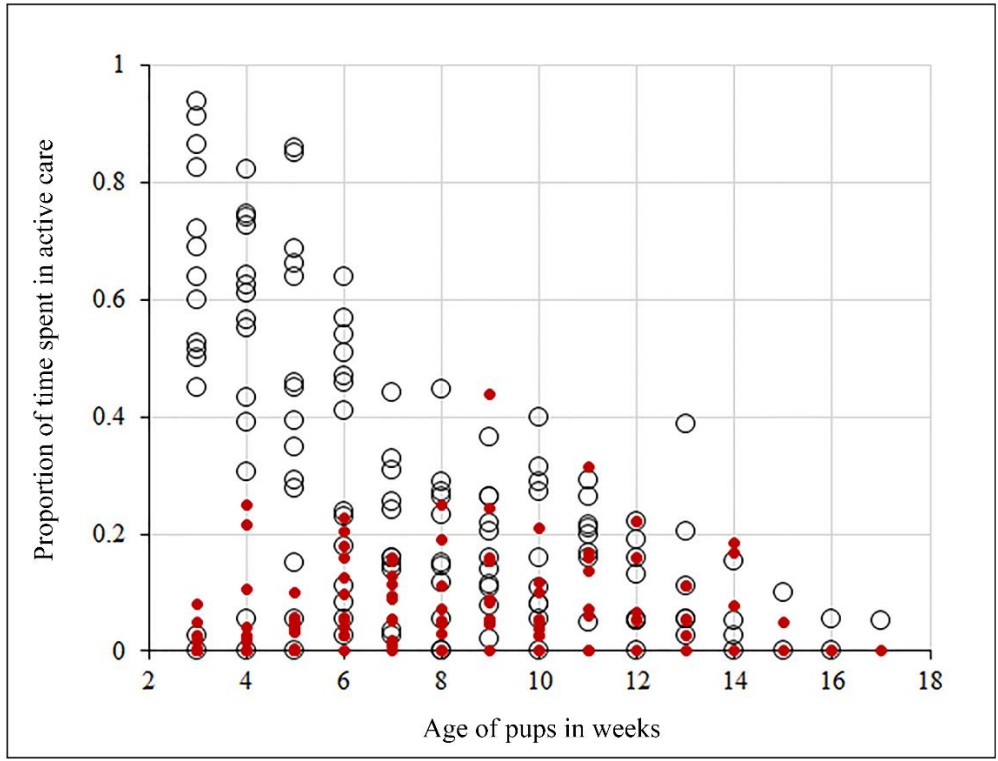

**Figure 2:** Scatter plot showing the proportion of time spent by the mother (empty circles) and allomothers (red circles) in active care over pup age.

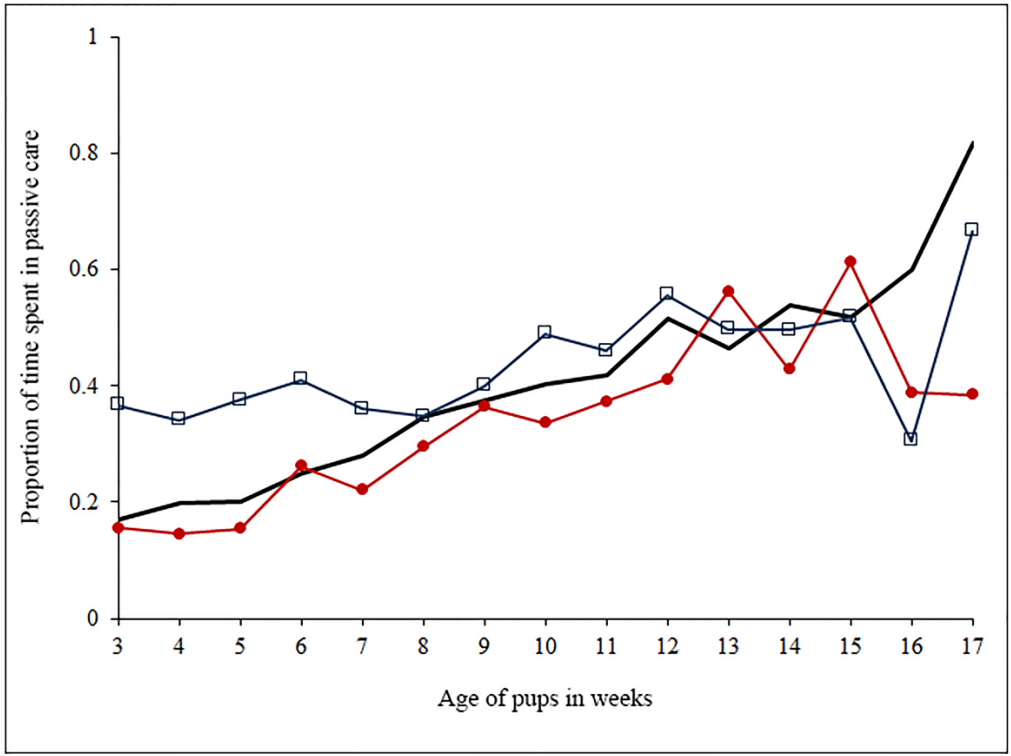

**Figure 3:** Line graph showing the proportion of time spent in passive care by the mothers (thick black line), putative fathers (thin black line with open squares) and allomothers (thin gray line with solid circles) over pup age.

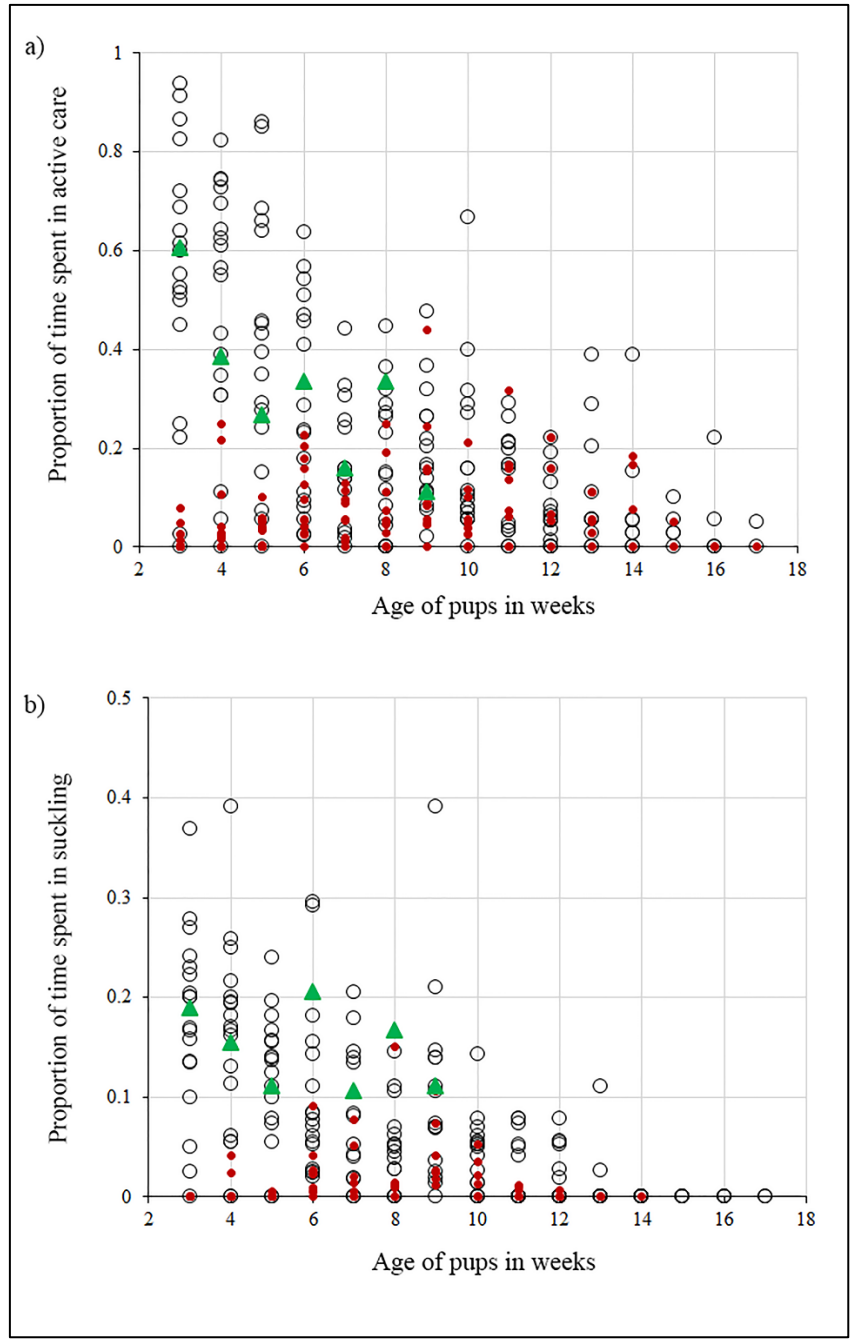

**Figure 4:** Scatterplots showing the similarity between the mothers and $PF_2$ grandmother over active care and suckling. a) Mothers (empty circles) and $PF_2$ grandmother (green triangles) spent comparable amounts of their time in active care unlike the allomothers (red circles). b) Proportion of time spent by $PF_2$ grandmother (green triangles) in suckling is comparable with the mothers (empty circles) but not with the allomothers (red circles).

**Supplementary information ESM 1**

**Total care**

Details of the linear mixed effect model that shows the effect of pup age and the current litter size on the proportion of time spent in "**total care**" by the putative fathers (PFs) and allomothers (AMs).

**Putative fathers (PFs)**

In order to check the effect of both the predictor variables i.e. pup age and the current litter size, we ran a "linear mixed effect model" incorporating the predictor variables as the "fixed effects" whereas the proportion of time spent by the PFs in total care was included in the model as the "response variable". We collected the data on male-allocare from 15 different dog groups, over a span of 5 years (2010-2015). Hence the identity of each mother-litter unit (fgr) and the year of data collection (fyr) were incorporated in the model as the "random effects". A Gaussian distribution was considered for the response variable in the model. We started with the full model, i.e., with all possible two-way interactions among the fixed effects.

**Variables used in the model:**

Response variable:

Proportion of time spent by the PFs in total care- **tcarepf**

Fixed effects:

Age of pups in weeks- **age**

Current litter size- **LS**

Random effects:

Group identity- **fgr**

Year of observation- **fyr**

Model: **modt.pf1<- lme (tcarepf~ age*LS, random = ~1|fgr/fyr)**

|  | Value | Std. Error | DF | t-value | p-value |
| --- | --- | --- | --- | --- | --- |
| (Intercept) | 0.4260810 | 0.12675825 | 125 | 3.361367 | 0.0010 |
| Age | 0.0169382 | 0.01088771 | 125 | 1.555715 | 0.1223 |
| LS | -0.0341776 | 0.03223116 | 125 | -1.060389 | 0.2910 |
| age * LS | 0.0034396 | 0.00320196 | 125 | 1.074206 | 0.2848 |

The two-way interaction showed no significant effect on the response variable and hence we reduced the model using standard protocol of backward selection method and ended up with the optimal model.

Model: **modt.pf2<- lme (tcarepf~ age + LS, random = ~1|fgr/fyr)**

|  | Value | Std. Error | DF | t-value | p-value |
| --- | --- | --- | --- | --- | --- |
| (Intercept) | 0.3345553 | 0.09900394 | 126 | 3.379212 | 0.0010 |
| Age | **0.0270875** | **0.00565348** | **126** | **4.791293** | **0.0000** |
| LS | -0.0062323 | 0.01902409 | 126 | -0.327598 | 0.7438 |

Pup age (age) showed significant effect on the total care shown by the PFs.

**Allomothers (AMs)**

In order to check the effect of both the predictor variables i.e. pup age and the current litter size, we ran a "linear mixed effect model" incorporating the predictor variables as the "fixed effects" whereas the proportion of time spent by the AMs in total care was included in the model as the "response variable". We collected the data on female-allocare from 15 different dog groups, over a span of 5 years (2010-2015). Hence the identity of each mother-litter unit (fgr) and the year of data collection (fyr) were incorporated in the model as the "random effects". A Gaussian distribution was considered for the response variable in the model. We started with the full model, i.e., with all possible two-way interactions among the fixed effects.

**Variables used in the model:**

Response variable:

Proportion of time spent by the AMs in total care- **tcaream**

Fixed effects:

Age of pups in weeks- **age**

Current litter size- **LS**

Random effects:

Group identity- **fgr**

Year of observation- **fyr**

Model: **modt.am1<- lme (tcaream~ age*LS, random = ~1|fgr/fyr)**

|  | Value | Std. Error | DF | t-value | p-value |
|---|---|---|---|---|---|
| (Intercept) | 0.16272215 | 0.13504054 | 120 | 1.2049874 | 0.2306 |
| Age | 0.03733670 | 0.01212582 | 120 | 3.0791071 | 0.0026 |
| LS | -0.01439017 | 0.03297950 | 120 | -0.4363368 | 0.6634 |
| age * LS | -0.00258630 | 0.00383986 | 120 | -0.6735394 | 0.5019 |

The two-way interaction showed no significant effect on the response variable and hence we reduced the model using standard protocol of backward selection method and ended up with the optimal model.

Model: **modt.am2<- lme (tcaream~ age + LS, random = ~1|fgr/fyr)**

|             | Value       | Std. Error  | DF  | t-value   | p-value |
|-------------|-------------|-------------|-----|-----------|---------|
| (Intercept) | 0.21143078  | 0.11482146  | 121 | 1.841387  | 0.0680  |
| age         | **0.03026155** | **0.00612711** | **121** | **4.938958** | **0.0000** |
| LS          | -0.03117045 | 0.02175889  | 121 | -1.432539 | 0.1546  |

Pup age (age) showed significant effect on the total care shown by the AMs.

## Supplementary information ESM 2

**Active care**

Details of the linear mixed effect model that shows the effect of pup age and the current litter size on the proportion of time spent in "**active care**" by the putative fathers (PFs) and allomothers (AMs).

**Putative fathers (PFs)**

In order to check the effect of both the predictor variables i.e. pup age and the current litter size, we ran a "linear mixed effect model" incorporating the predictor variables as the "fixed effects" and the proportion of time spent by the PFs in active care as the "response variable". We collected the data on male-allocare from 15 different dog groups, over a span of 5 years (2010-2015). Hence the identity of each mother-litter unit (fgr) and the year of data collection (fyr) were incorporated in the model as the "random effects". A Gaussian distribution was considered for the response variable in the model. We started with the full model, i.e., with all possible two-way interactions among the fixed effects.

**Variables used in the model:**

Response variable:

Proportion of time spent by the AMs in active care- **acarepf**

Fixed effects:

Age of pups in weeks- **age**

Current litter size- **LS**

Random effects:

Group identity- **fgr**

Year of observation- **fyr**

Model: **moda.pf1<- lme (acarepf~ age*LS, random = ~1|fgr/fyr)**

|            | Value       | Std. Error | DF  | t-value    | p-value |
|------------|-------------|------------|-----|------------|---------|
| (Intercept)| 0.09804374  | 0.06760597 | 125 | 1.4502230  | 0.1495  |
| age        | -0.00068232 | 0.00519360 | 125 | -0.1313768 | 0.8957  |
| LS         | -0.01430338 | 0.01622166 | 125 | -0.8817458 | 0.3796  |
| age * LS   | 0.00228893  | 0.00148019 | 125 | 1.5463719  | 0.1245  |

The two-way interaction showed no significant effect on the response variable and hence we reduced the model using standard protocol of backward selection method and ended up with the optimal model.

Model: **moda.pf2<- lme (acarepf~ age + LS, random = ~1|fgr/fyr)**

|            | Value       | Std. Error | DF  | t-value   | p-value |
|------------|-------------|------------|-----|-----------|---------|
| (Intercept)| 0.03572216  | 0.05493874 | 126 | 0.6502181 | 0.5167  |
| age        | **0.00612957** | **0.00277573** | **126** | **2.2082712** | **0.0290** |
| LS         | 0.00520547  | 0.01034108 | 126 | 0.5033780 | 0.6156  |

Pup age (age) showed significant effect on the active care shown by the PFs.

**Allomothers (AMs)**

In order to check the effect of both the predictor variables i.e. pup age and the current litter size, we ran a "linear mixed effect model" incorporating the predictor variables as the "fixed effects" and the proportion of time spent by the AMs in active care as the "response variable". We collected the data on female-allocare from 15 different dog groups, over a span of 5 years (2010-

2015). Hence the identity of each mother-litter unit (fgr) and the year of data collection (fyr) were incorporated in the model as the "random effects". A Gaussian distribution was considered for the response variable in the model. We started with the full model, i.e., with all possible two-way interactions among the fixed effects.

**Variables used in the model:**

Response variable:

Proportion of time spent by the AMs in active care- **acaream**

Fixed effects:

Age of pups in weeks- **age**

Current litter size- **LS**

Random effects:

Group identity- **fgr**

Year of observation- **fyr**

Model: **moda.am1<- lme (acaream~ age*LS, random = ~1|fgr/fyr)**

|  | Value | Std. Error | DF | t-value | p-value |
| --- | --- | --- | --- | --- | --- |
| (Intercept) | 0.023374113 | 0.04656811 | 120 | 0.5019339 | 0.6166 |
| age | 0.006505368 | 0.00407203 | 120 | 1.5975733 | 0.1128 |
| LS | 0.006515888 | 0.01036865 | 120 | 0.6284221 | 0.5309 |
| age * LS | -0.000940014 | 0.00129656 | 120 | -0.7250054 | 0.4699 |

The two-way interaction showed no significant effect on the response variable and hence we reduced the model using standard protocol of backward selection method and ended up with the optimal model.

Model: **moda.am2<- lme (acaream~ age + LS, random = ~1|fgr/fyr)**

|             | Value      | Std. Error | DF  | t-value   | p-value |
|-------------|------------|------------|-----|-----------|---------|
| (Intercept) | 0.04242796 | 0.03867422 | 121 | 1.0970607 | 0.0680  |
| age         | **0.00388757** | **0.00187833** | **121** | **2.0696973** | **0.0406** |
| LS          | 0.00007415 | 0.00525762 | 121 | 0.0141041 | 0.9888  |

Pup age (age) showed significant effect on the active care shown by the AMs.

# Supplementary information ESM 3

## Passive care

Details of the linear mixed effect model that shows the effect of pup age and their current litter size on the proportion of time spent in "**passive care**" by the putative fathers (PFs) and allomothers (AMs).

## Putative fathers (PFs)

In order to check the effect of both the predictor variables i.e. pup age and the current litter size, we ran a "linear mixed effect model" incorporating the predictor variables as the "fixed effects" and the proportion of time spent by the PFs in passive care as the "response variable". We collected the data on male-allocare from 15 different dog groups, over a span of 5 years (2010-2015). Hence the identity of each mother-litter unit (fgr) and the year of data collection (fyr) were incorporated in the model as the "random effects". A Gaussian distribution was considered for the response variable in the model. We started with the full model, i.e., with all possible two-way interactions among the fixed effects.

## Variables used in the model:

Response variable:

Proportion of time spent by the PFs in passive care- **pcarepf**

Fixed effects:

Age of pups in weeks- **age**

Current litter size- **LS**

Random effects:

Group identity- **fgr**

Year of observation- **fyr**

Model: **modp.pf1<- lme (pcarepf~ age*LS, random = ~1|fgr/fyr)**

|  | Value | Std. Error | DF | t-value | p-value |
|---|---|---|---|---|---|
| (Intercept) | 0.23758554 | 0.14014505 | 125 | 1.6952831 | 0.0925 |
| age | 0.02203842 | 0.01045623 | 125 | 2.1076839 | 0.0371 |
| LS | -0.00255958 | 0.03226917 | 125 | -0.0793197 | 0.9369 |
| age * LS | 0.00082516 | 0.00299011 | 125 | 0.2759621 | 0.7830 |

The two-way interaction showed no significant effect on the response variable and hence we reduced the model using standard protocol of backward selection method and ended up with the optimal model.

Model: **modp.pf2<- lme (pcarepf~ age + LS, random = ~1|fgr/fyr)**

|  | Value | Std. Error | DF | t-value | p-value |
|---|---|---|---|---|---|
| (Intercept) | 0.21470119 | 0.11486552 | 126 | 1.869153 | 0.0639 |
| age | **0.02450343** | **0.00550545** | **126** | **4.450761** | **0.0000** |
| LS | 0.00442425 | 0.02004886 | 126 | 0.220673 | 0.8257 |

Pup age (age) shown significant effect on the passive care shown by the PFs.

**Allomothers (AMs)**

In order to check the effect of both the predictor variables i.e. pup age and the current litter size, we ran a "linear mixed effect model" incorporating the predictor variables as the "fixed effects" and the proportion of time spent by the AMs in passive care as the "response variable". We collected the data on female-allocare from 15 different dog groups, over a span of 5 years (2010-

2015). Hence the identity of each mother-litter unit (fgr) and the year of data collection (fyr) were incorporated in the model as the "random effects". A Gaussian distribution was considered for the response variable in the model. We started with the full model, i.e., with all possible two-way interactions among the fixed effects.

**Variables used in the model:**

Response variable:

Proportion of time spent by the AMs in passive care- **pcaream**

Fixed effects:

Age of pups in weeks- **age**

Current litter size- **LS**

Random effects:

Group identity- **fgr**

Year of observation- **fyr**

Model: **modp.am1<- lme (pcaream~ age*LS, random = ~1|fgr/fyr)**

|  | Value | Std. Error | DF | t-value | p-value |
| --- | --- | --- | --- | --- | --- |
| (Intercept) | 0.14133616 | 0.12035277 | 120 | 1.1743490 | 0.2426 |
| age | 0.03006444 | 0.01102891 | 120 | 2.7259656 | 0.0074 |
| LS | -0.02179411 | 0.02992098 | 120 | -0.7283891 | 0.4678 |
| age * LS | -0.00137052 | 0.00349478 | 120 | -0.3921627 | 0.6956 |

The two-way interaction showed no significant effect on the response variable and hence we reduced the model using standard protocol of backward selection method and ended up with the optimal model.

Model: **modp.am2<- lme (pcaream~ age + LS, random = ~1|fgr/fyr)**

|             | Value       | Std. Error  | DF  | t-value    | p-value |
|-------------|-------------|-------------|-----|------------|---------|
| (Intercept) | 0.16712635  | 0.10099926  | 121 | 1.654728   | 0.1006  |
| age         | **0.02631783** | **0.00554810** | **121** | **4.743573** | **0.0000** |
| LS          | -0.03068309 | 0.01961587  | 121 | -1.564197  | 0.1204  |

Pup age (age) showed significant effect on the passive care shown by the AMs.

**Supplementary information ESM 4**

Details of the ANOVA- Post-Hoc Tukey tests showing the comparisons among the mothers, $PF_2$ grandmothers and allomothers over active care and suckling.

**Comparisons among the proportion of time spent in active care by the mothers, allomothers and $PF_2$ grandmother:**

| Analysis of Variance | | | | |
|---|---|---|---|---|
| Source | Df | Mean Sq. | F | Prob. |
| Model | 2 | 0.148 | 8.424 | 0.003 |

| Post Hoc tests | | | | | | |
|---|---|---|---|---|---|---|
| Test | Group 1 | Group 2 | Mean Diff. | SE | q | Prob. |
| Tukey | Mothers | Allomothers | 0.256 | 0.050 | 5.106 | **0.005** |
| | | $PF_2$ grandmother | 0.008 | 0.050 | 0.162 | 0.993 |
| | Allomothers | $PF_2$ grandmother | -0.247 | 0.050 | 4.944 | **0.007** |

**Comparisons among the proportion of time spent in suckling by the mothers, allomothers and $PF_2$ grandmother:**

| Analysis of Variance | | | | |
|---|---|---|---|---|
| Source | Df | Mean Sq. | F | Prob. |
| Model | 2 | 0.032 | 24.831 | 0.000 |

| Post Hoc tests | | | | | | |
|---|---|---|---|---|---|---|
| Test | Group 1 | Group 2 | Mean Diff. | SE | q | Prob. |
| Tukey | Mothers | Allomothers | 0.093 | 0.014 | 6.900 | **0.000** |
| | | $PF_2$ grandmother | -0.038 | 0.014 | 2.778 | 0.150 |
| | Allomothers | $PF_2$ grandmother | -0.131 | 0.014 | 9.678 | **0.000** |